\documentclass[superscriptaddress,floatfix,twocolumn,prl]{revtex4}
\usepackage{graphicx}
\usepackage[usenames]{color}
\pdfoutput=1 
\begin{document}

\title{Probing ultrafast symmetry breaking in photo-stimulated matter}

\author{Daniele Fausti\footnote{present address: Max Plank Research Group for Structural dynamics,
CFEL, Desy, Notkerstrasse 85, 22607 Hamburg, Germany, email:
daniele.fausti@desy.de}} \affiliation{Zernike Institute for Advanced
Materials, University of Groningen, 9747 AG Groningen, The
Netherlands.}

\author{Oleg V. Misochko}
\affiliation{Institute of Solid State Physics, Russian Academy of
Sciences, 142432 Chernogolovka, Russia.}

\author{Paul H.M. van Loosdrecht\footnote{email: P.H.M.van.Loosdrecht@rug.nl}}
\affiliation{Zernike Institute for Advanced Materials, University of
Groningen, 9747 AG Groningen, The Netherlands.}

\begin{abstract}

Picosecond Raman scattering is used to study the photo-induced ultrafast
dynamics in Peierls distorted Antimony. We find evidence for an
ultrafast non-thermal reversible structural phase transition. Most
surprisingly, we find evidence that this transition
evolves toward a lower symmetry, in contrast to the commonly
accepted rhombohedral-to-simple cubic transition path. Our study
demonstrates the feasibility of ultrafast Raman scattering symmetry
analysis of photo-induced non-thermal transient phases.
\end{abstract}

\maketitle

Controlling the state of a material through light
irradiation and thereby obtaining transient highly-off-equilibrium
phases, is one of the intriguing achievements in condensed matter science 
of the last decade.\cite{Cho03,Cav04,rin07} Progress in this field has been
boosted by the easy availability of extremely short light pulses
(10$^{-14}$ s), kindling the hope of controlling matter on ultrafast
time-scales, {\em i.e.} on timescales faster than the characteristic
thermodynamical timescale which limit the speed of current phase change 
media based devices. 
The main emerging limitation in bidirectional ultrafast
optical switching is that by and large the photo-induced phase
transitions reported to date are low to high symmetry transitions.
Rare are the examples of photo-induced high to low symmetry
transitions necessary to complete the ultrafast bi-directional switching
cycle and controversial the possibility of photo-inducing a
high to low symmetry phase transition on ultrafast 
timescales\cite{Col03,maz08},
while the low-to-high symmetry phase transition can occur through direct
coupling of the the light field (carrying little momentum) to
crystal excitations, the reduction of the symmetry can only arise
due to a cooperative effect leading to self-organized long-range
order which is usually limited to occur on 'thermodynamic'
timescales. In contrast to this, our findings from ultrafast
experiments on the A7 metals demonstrate the possibility of inducing
a reversible photo-induced symmetry lowering on timescales
surpassing the time needed to reach thermodynamical equilibrium.

\begin{figure}[htb]
\includegraphics[width=8cm]{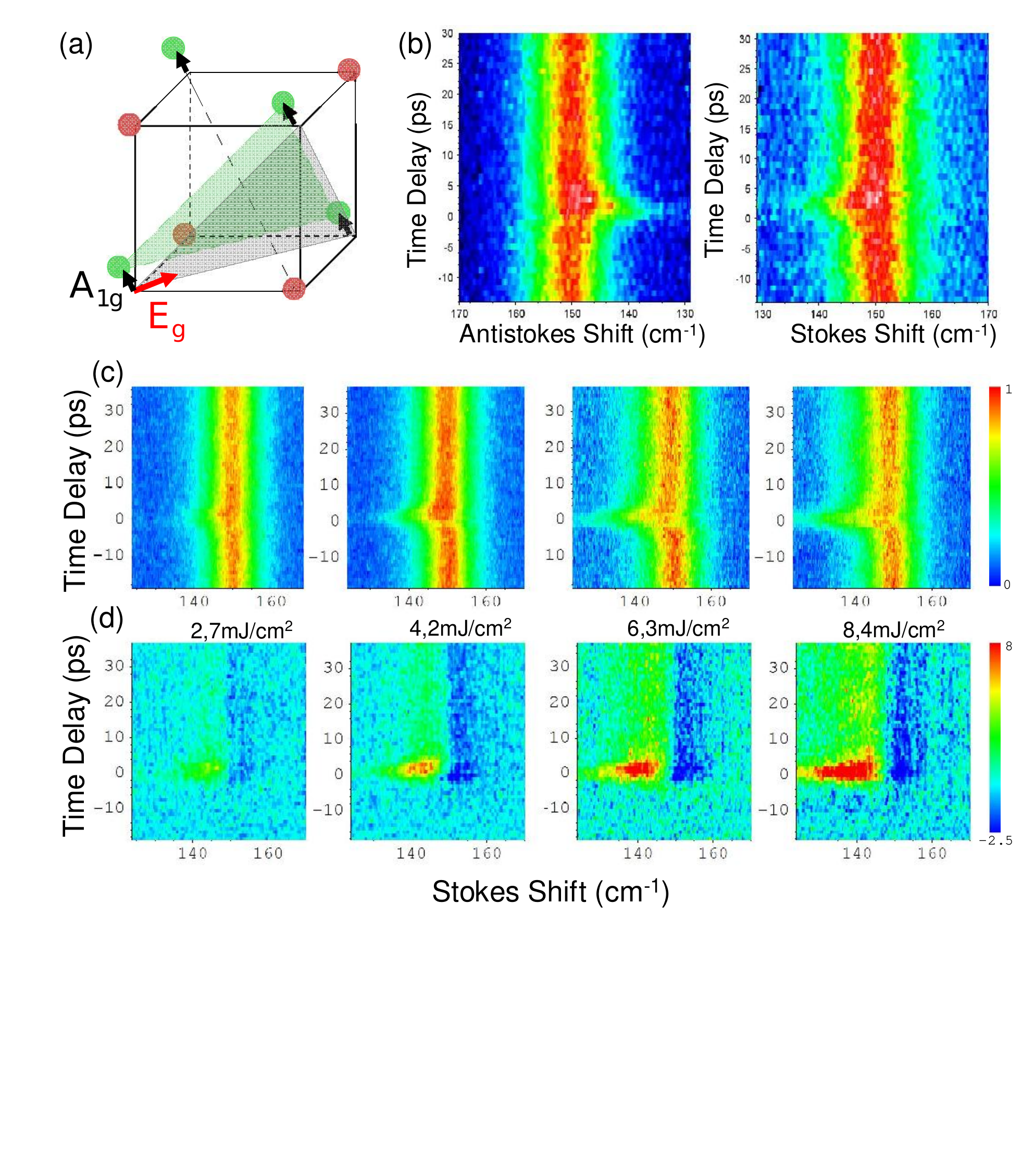}
\caption{ 
{\bf (a)}
Equilibrium structure of the A7 semimetals (space group
R$\overline{3}m$). The A$_{1g}$ vibrational mode corresponds 
to a modulation of the distance between the (111) planes,
while the E$_{g}$ mode corresponds to a sliding of adjacent (111) planes.
{\bf (b)} 
False color plot of the time 
resolved Stokes (left) and anti-Stokes (right) Raman data 
for excitation density of 4.6mJ/cm$^{2}$. 
{\bf (c)}
False color plot of the pump-probe Raman data for various pump
intensities.
{\bf (d)} Optically induced changes in the Raman spectra
obtained by the subtraction of the negative time response from the
data in (c).} \label{fig2}
\end{figure}

The group V semi-metals like Bismuth and Antimony have served as a
playground for studying interactions between ultrafast light pulses
and absorbing matter.\cite{Gar97,Has98,DeCam01,Mis2004} The strong
coupling between the structural and electronic degrees of freedom in
these materials allowed for the first pioneering studies in the
late-eighties, early-nineties on ''coherent phonon'' generation in
absorbing materials, visualizing the real time behavior of optical
phonons.\cite{tho84,che90,zei92} More recently the structural and
electronic dynamics following laser irradiation in Bi and Sb single
crystals have been studied in great detail, and the possibility of
inducing a ''non-thermal'' phase transition to a simple cubic phase
has been discussed on the base of ab-initio
calculations.\cite{mur05,zij06,fri07,mur07}
The structure of the A7 compounds (sketched in Fig.\ref{fig2}a) may
be described as a distorted simple cubic structure, where the (111)
planes of atoms have an alternating displacement along the [111]
direction. This structural peculiarity of the semi metals Bi, Sb,
and As has been widely discussed in the past\cite{Boy60,Edel76,iwa97}, and
originates from a strong electron phonon coupling. In one dimension,
this type of distortion is the well known Peierls
distortion.\cite{jon34, Peierls} 

The physics behind the expected photo-induced effects in A7 metals
can be sketched in a simple way. The photo-excitation of 
valence band electrons increases the electron
density in the conduction band and, as a consequence, it reduces the
energy gain of the Peierls distortion. Eventually this renders the
Peierls distortion unstable, and a phase transition should occur to
the undistorted phase on a time-scale faster than the time required
for electron-phonon thermalization. The ab-initio calculations meantioned
earlier indicated that the cubic phase should be reached for a critical
excitation density of 2.7 electrons per hundred
ions.\cite{mur05,mur07} Various experimental studies have tried to
reach this excitation limit and to detect an optically induced phase
transition in Bismuth and Antimony single crystals making use of a
variety of techniques, including time-resolved
reflectivity,\cite{bos08} x-ray diffraction,\cite{fri07,Sok03} and
UV absorption experiments.\cite{pap08} In spite of the experimental
efforts devoted, no evidence of an optically induced reversible
phase transition in the A7 semimetals has been reported so far under
any of the explored excitation conditions.

The current work presents results of an ultrafast time resolved Raman 
spectroscopy study of Antimony single crystals
\footnote{We have also performed a study of Bismuth,
which results completely analogous to those reported in this article
on Antimony}. This technique is sensitive to transient changes in
the crystal structure through transient changes in the spontaneous
vibrational Raman response. In addition to this, a comparison
between the Stokes and anti-Stokes scattering intensity allows to
distinguish the dynamics induced by lattice heating effects from the
non-thermal ones arising from the electronic screening of the
crystal potential. The experiments reveal two distinct dynamics: A
fast non-thermal response occurring in the first few picoseconds
after irradiation, and a slower thermal one which lasts for more
than 100ps. The short time response evidences the existence of an
induced transient state, even though the nature of this state
deviates from the expected simple cubic phase.

The experiments were performed using a 80 MHz picosecond Ti:Sapphire
laser (MIRA 900, wavelength 800 nm, pulse duration 1.7 ps) as
excitation laser, which was coupled using a custom designed
pump-probe scheme, to the microscope of a standard charge coupled
device equipped Raman spectrometer (T64000, Jobin Yvon). All
experiments have been performed at ambient conditions in a
controlled Argon environment.

Fig.\ref{fig2}(b) shows the time evolution of the A$_{1g}$ mode in
Antimony after excitation with a moderately low pump power density
(4.6 mJ/cm$^2$). The right and left panels display a false color
plot of the Stokes and anti-Stokes response, respectively. In the
first few picoseconds after excitation, the A$_{1g}$ mode shows a
rapid and appreciable softening toward the low energy side, with a
subsequent relaxation, in roughly 10 ps, to a transient spectrum
which is close to the equilibrium one. This transient state slowly
relaxes to the equilibrium situation, with a decay time exceeding
100 ps. The dynamical response is demonstrated more clearly in 
Fig.~\ref{fig2}(c) and (d), which display in a false color plot the
direct (b) and differential (c) transient Stokes response for four
different excitation densities.

\begin{figure}[htb]
\includegraphics[width=8cm]{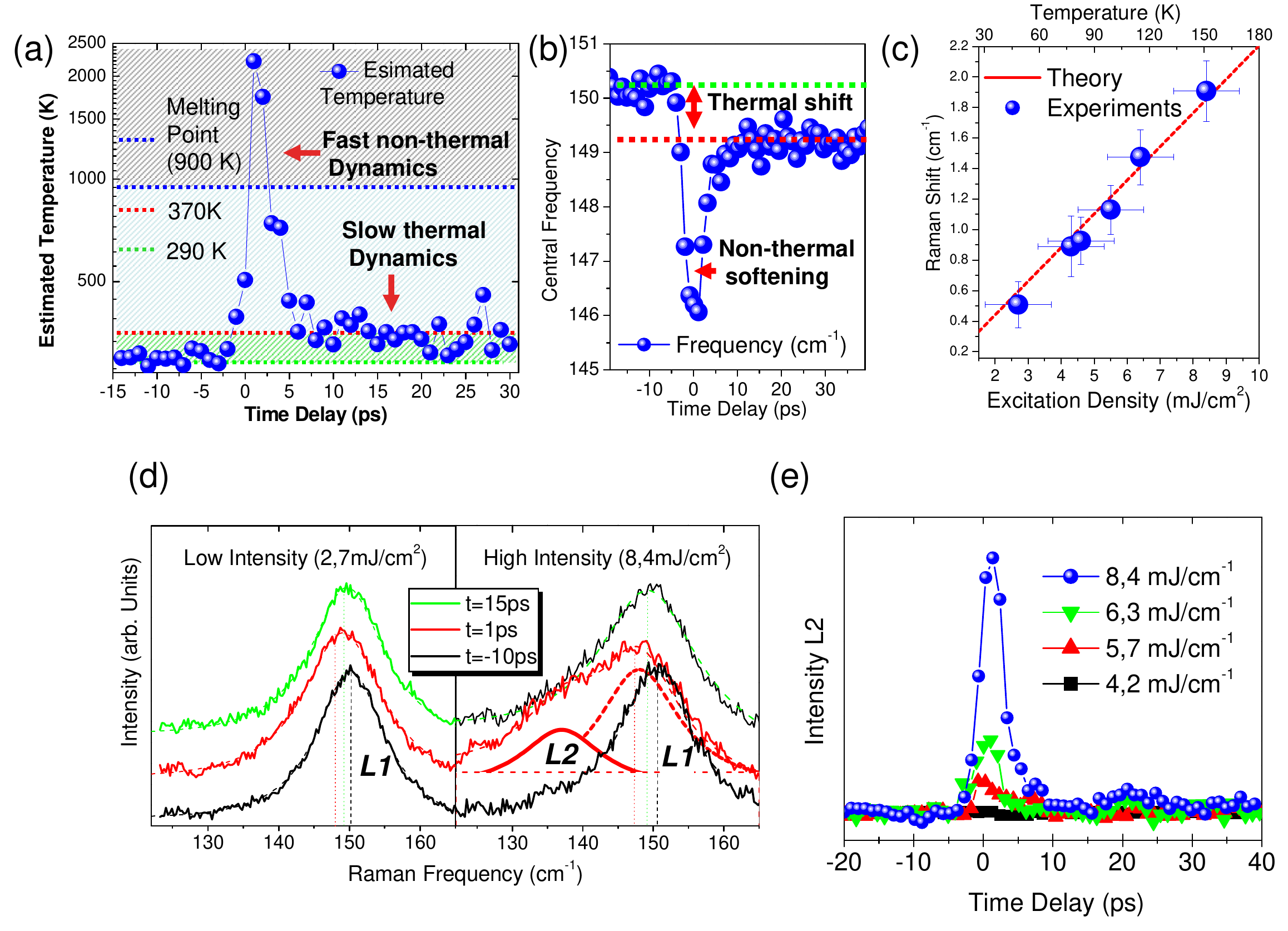}
\caption{ 
{\bf (a)} 
Time evolution of the A$_{1g}$ phonon temperature after excitation
with 4.6mJ/cm$^{2}$. The blue dashed line indicates the thermodynamical 
melting temperature. The average long time laser induced heating is indicated 
by the red dashed line.
{\bf (b)}
Time evolution of the central frequency of the Stokes line for a pump excitation of 4.6mJ/cm$^{2}$. 
{\bf (c)} 
Thermal frequency shift of the $A_{1g}$ phonon at 20 picoseconds after pump excitation versus the pump
excitation density and the expected temperature variation. 
The dashed red line shows the temperature dependence of the frequency
shift obtained from continuous wave experiments. 
{\bf (d)} 
Raman response for the unperturbed system (t=-10 ps, black curve),
for the ultrafast non-thermal response (t=1ps, red curve) and for
the thermal response (t=20 ps, green curve). The left and right 
panels depicts Raman spectra in the low and high excitation density
limit, respectively. 
{\bf (e)} 
Time traces of the intensity of
the new phonon line for various excitation densities.} \label{fig3}
\end{figure}

The ratio of the intensity of the Stokes and anti-Stokes spectra
allow extraction of the non-equilibrium phonon temperature as a
function of time. The evolution of the phonon temperature after the
pump excitation with a density of 4.6 mJ/cm$^2$ is depicted in
Fig.\ref{fig3}(a). In the first 3 ps the phonon temperature raises
to more then 2000 K. Clearly this does not correspond to a
thermodynamic temperature raise, but merely reflects the
non-equilibrium phonon occupation which is boosting the anti-Stokes
response. On the same timescale, however, also the frequency of the
A$_{1g}$ mode softens considerably (see Fig.\ref{fig3}(b)) to values
which are not achievable under equilibrium conditions (at
atmospheric pressure\cite{deg07}). This indicates that not only population
effects, but also electronic screening occurs on this timescale. We
will return to discuss the fast response later on after having
discussed the nature of the spectral changes at large positive times
($t>10$~ps).

Equilibrium between the electronic and the lattice temperature is
reached in about 10 picoseconds after pump irradiation. Based on the
equilibrium thermodynamical and optical properties of Antimony, one
can estimate the expected temperature raise resulting from the photo
excitation. As an example, based on the reflectivity, optical
penetration depth, and heat capacity of Antimony, on expects that
excitation with a power density of 4.6 mJ/cm$^{-2}$ results in a
temperature raise of approximately 80K. This is indeed what is
observed experimentally, as is indicated by the red dashed line in
Fig.\ref{fig3}(a). The temperature raise estimated from the
intensity of the anti-Stokes signals is confirmed by the frequency
shift of the A$_{1g}$ phonon, as is indicated by the red dashed line in
Fig.\ref{fig3}(b) which indicates the expected frequency for Sb at
$T=370$~K\cite{Hon77}. The residual frequency shift observed at
times larger than 10 ps confirms the completed thermalization
between the electronic and lattice subsystems. This is further
illustrated in Fig.\ref{fig3}(c), which compares frequency of the
A$_{1g}$ mode as a function of the measured phonon temperature at 20
ps after excitation (symbols) with the result of continuous wave
Raman measurements as a function of the thermodynamic
temperature\cite{Hon77} (red line).
The good agreement of the experimental and estimated temperature (at
long times $>$10 ps) indicates that diffusive heat transport by
photo-excited electrons is less efficient than energy relaxation to
the lattice. Nearly all the energy dumped by the optical pulse and
adsorbed by the electronic subsystem results in local lattice heating
within 10 ps. Vibrational and electronic heat diffusion only play a
significant role at later times when the system slowly relaxes back
to the initial state (t$>$100ps).

Now we turn to back to the ultrafast response. Two features clearly
evidence the non-thermal nature of the processes occurring at early
times following laser irradiation. Firstly the calculated phonon 
temperature in the first 10 picoseconds reaches values up to 2200 K,
well above the equilibrium melting temperature of Antimony
(indicated by the blue dashed line in Fig.\ref{fig3}(a)). Secondly,
the frequency of the A$_{1g}$ mode in the first 10 picoseconds
reaches values lower than those measured under equilibrium
conditions at any temperature up to the melting point. The observed
non-thermal phonon softening is comparable to the one measured in
the time-domain with similar electronic excitation densities,
clearly demonstrating that the electronic screening of the crystal
potential induces a large phonon softening and that anharmonicity
only plays a minor role, if any.

\begin{figure}[htb]
\includegraphics[width=8cm]{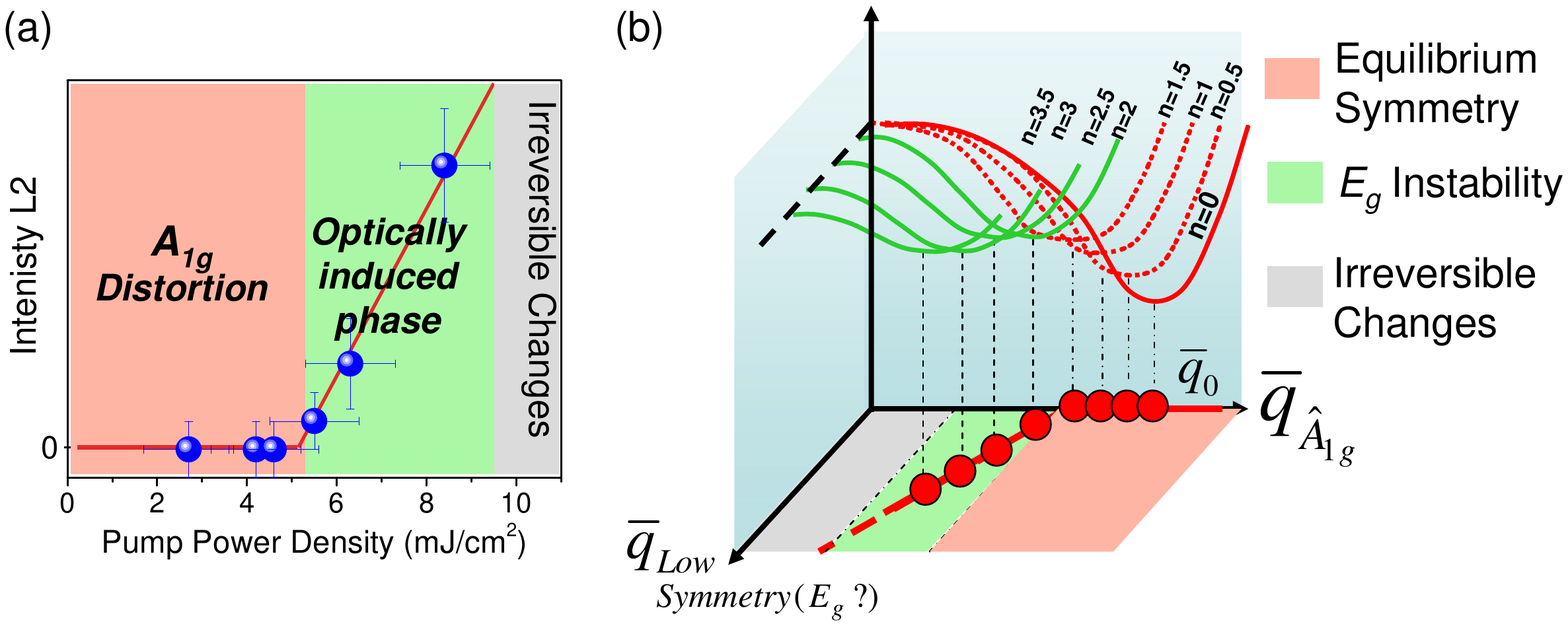}
\caption{
{\bf (a)} 
Maximum intensity of the new vibrational mode (L2) of the induced phase as a
function of pump power density.  
{\bf (b)} Sketch of the induced non-linear dynamics in the A7 metals. 
The curves represent cuts of the free energy for
a distortion along the A$_{1g}$ direction. For low density
electronic excitation the minimum of the free energy stays on the
fully symmetric (A$_{1g}$) direction thereby preserving the A7
symmetry. Excitation densities exceeding 2 electron per 100 atoms (5
mJ/cm$^{2}$) lead to a symmetry breaking due to
displacement along a low symmetry direction (possibly $E_{g}$).}
\label{fig4}
\end{figure}

One of the most striking observations it that the line shape for the
first few picoseconds  after excitation is substantially different
for low and high excitation densities. This is illustrated in
Fig.\ref{fig3}(d), where the upper and lower panels show  the Raman
response in the region of the A$_{1g}$ mode at different times for
low (2.7mJ/cm$^{2}$) and high (8.4mJ/cm$^{2}$) excitation density,
respectively. Both before arrival of the pump pulse (black traces)
as well as at late times (green traces) the phonon response shows
the standard Lorentian distribution (L1) for both excitation
densities. In contrast, the early time responses (red traces) differ
drastically. For moderately low power density ($<$5 mJ/cm$^{2}$) the
A$_{1g}$ phonon shifts to lower frequency, keeping a Lorentian shape
response with approximately the same linewidth as measured before
pumping, indicating that the A7 structure is retained. This is no
longer true for power densities exceeding ~5mJ/cm$^{2}$. In this
case the early time Raman spectrum shows, apart from the normal
response (L1) the appearance of a new shoulder (L2) at the
low energy side. This strongly suggests that at high excitations
density the symmetry of antimony has changed substantially in the
first few picoseconds, {\it i.e.} that an optical phase transition
has occurred. Further strong evidence for this comes from the
threshold behavior observed in the activation of the L2 mode as
demonstrated in Fig.~\ref{fig4}(a). The mode is only observed for
excitation densities exceeding 5 mJ/cm$^2$, and shows a linear
increase of intensity upon further increasing the excitation
density.

A two component fit of the high excitation density Raman spectra at
different delay times reveals that the additional phonon mode,
unlike the L1 mode, has a time independent frequency. Moreover, the
frequency of the new mode is $\simeq$10\% lower than the frequency
of the equilibrium A$_{1g}$ mode, and $\simeq$20\% higher than the
E$_g$ mode. This latter mode, which is active in a different
scattering geometry, could be activated by optically induced
disorder. These observations demonstrate that this new mode is
indeed not originating from the A7 structure. Finally we note that
the optically induced phase has a relatively short lifetime ($\sim$5
ps) as shown by the time dependence of the intensity plotted in
Fig.\ref{fig3}(e).

The question now arises what the symmetry of the new optically
induced phase is. It is clear that the long expected phase
transition to the high symmetry cubic phase can be excluded.
For the simple cubic symmetry one expects no optical phonon mode at
all, in striking contrast with the presence of the observed
additional mode. Moreover, one would expect a complete softening of
the A$_{1g}$ mode which is clearly not observed. No anti-correlation
between the intensities of the L1 and L2 modes has been observed.
This makes it unlikely that the observed new mode is due to a new
phase which coexists with the A7 structure. We therefore conclude
that both observed modes originate from the optically induced phase,
and that this phase has a lower symmetry than the A7 structure. One
intriguing possibility is that the structural change is not due to a
softening of the A$_{1g}$ mode, but rather due to a E$_g$
distortion, {\em i.e.} due to an alternating displacement of the
(111) planes along a direction perpendicular to the [111] direction
(indicated in Fig.\ref{fig2}a). Such a distortion would most likely
lead to a monoclinic C$2/m$ structure, with indeed an activation of
additional modes in the Raman spectrum.\footnote{The dispersion
relations of the E$_{g}$ modes in Antimony single
crystal\cite{Sha71} show that a doubling of the unit cell would give
an additional zone center mode at a frequency close to the observed
one.}

Fig.\ref{fig4} (b) summarizes the mechanism suggested for the
optically-induced phase transition in the A7 metals. The excitation
of electrons into the conduction bands reduces the energy gain of
the Peierls distortion forcing the ions to move towards the cubic
undistorted symmetry (red area in Fig.\ref{fig4} (a) and (b)). As
the excitation density is increased the alternating distortion of
the (111) planes along the A$_{1g}$ direction reduces while
retaining the overall rhombohedral symmetry. For excitation
densities exceeding $\sim$5mJ/cm$^{2}$ an instability (possibly of
E$_{g}$ nature) sets  in (green dashed area in Fig.\ref{fig4} (a)
and (b)) which reduces the A7 symmetry without ever reaching the
high symmetry cubic phase. We believe that in our configuration we
can explore this region of the phase diagram thanks to the fact that
our measurements do not require the non-adiabatic coherent
excitation of the fully symmetric A$_{1g}$ mode for detection. The
collectiveness of the coherent excitation of a fully symmetric mode
could increase the energy cost of displacing the ions towards a
lower symmetry location and would indicate why such a transient
response has never been observed with different techniques.

In conclusion we demonstrated the feasibility of time resolved Raman
studies to unravel the interplay between electronic and structural
degrees of freedom. Insight into the thermalization processes
between electrons and ions have been obtained from the measured
transient Raman spectra, showing the thermal nature of the crystal
response for times larger that 10 picoseconds after the pump
excitation, and demonstrating that electronic energy diffusion plays
a minor rule in the transient dynamics for the A7 metals. In the
non-thermal region (t$<$10 picoseconds) we demonstrate that the
electronic screening of the crystal potential can induce a large
phonon softening ruling out that anharmonicity plays an important
role. Maybe the most important result of the present work is the
observation of an ultrafast optically induced phase transition
toward a non-thermal low symmetry phase for excitations exceeding ~2
electron per 100 ions. Ultrafast symmetry lowering transitions are
important for ultrafast bidirectional switching and contrasts most
of the observed ultrafast phase transitions to date.

{\bf Acknowledgements}. We sincerely thank Ben Hesp, Arjen F.
Kamp and Foppe de Haan for their help in executing the experiments 
and analyzing the data. We also sincerely thank Prof.
Cavallieri and Dr. Mostovoy for fruitful discussions and are grateful to Dr. Giannetti, Dr. Colombi and Dr.
Zaffalon for the critical reading of the manuscript.\\
This work is part of the research programme of the 
'Stichting voor Fundamenteel Onderzoek der Materie (FOM)', which is financially supported by the 
'Nederlandse Organisatie voor Wetenschappelijk Onderzoek (NWO)'.


\end{document}